# Steady state theory of current transfer

Vered Ben Moshe and Abraham Nitzan
School of Chemistry, Tel Aviv University, Tel Aviv 69978, Israel
Spiros S. Skourtis
Department of Physics, University of Cyprus, Nicosia 1678, Cyprus
David Beratan
Department of Chemistry, Duke University, Durham, NC 27708 USA

## Abstract

Current transfer is defined as a charge transfer process where the transferred charge carries information about its original motion. We have recently suggested that such transfer causes the asymmetry observed in electron transfer induced by circularly polarized light through helical wires. This paper presents the steady state theory of current transfer within a tight binding model of coupled wires systems. The efficiency of current transfer is quantified in terms of the calculated asymmetry in the system response to a steady current imposed on one of the wires, with respect to the imposed current direction.

## 1. Introduction

*Current transfer* is defined as a charge transfer transition chracterized by relocation of both charge and its momentum. In a recent paper[1] we have proposed a tight binding charge transfer model for recent observations[2,3] that indicate that photo-electron transfer induced by circularly polarized light through helical molecular bridges depends on the relative handedness of the bridge helicity and on the optical circular polarization. Another recent example of current transfer in photoemission is provided by Ref. [4], in which the signature of a biased linear momentum distribution created on Cu (100) surface is observed in the angular distribution of the photoemitted current. Our rationalization of the experimental results of Refs. [2,3] was based on the assumption (supported by theoretical analysis[5-8] that excitation by circularly polarized light can create a circular electronic current in the absorbing molecule and that chiral control of transmission of these currents results from a coupling scheme associated with atoms proximity. Figure 1 illustrates this idea.

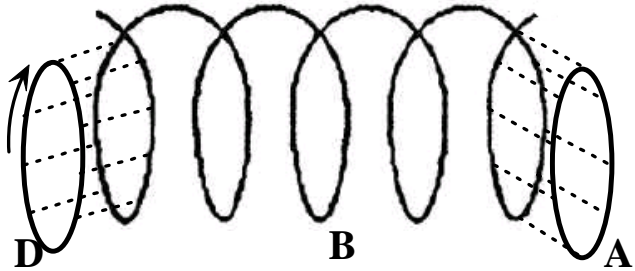

Figure 1. A circular current in the donor (D) ring is transferred to an acceptor A through a helical bridge. The dominant intermolecular coupling, illustrated by dotted lines connecting nearest atoms belonging to neighboring molecules can coherently transmit directional information. The clockwise circular current indicated by the arrow on the donor will be transmitted through the helical bridge shown more readily than a circular current in the opposite direction.

It should be clear from Fig.1 that the current transfer phenomenon originates from the coupling scheme, which results here from proximity of circular molecular structures. Simpler structures that show the same physical behavior are displayed in Figure 2, where each structure corresponds to a tight binding Hamiltonian with nearest neighbor coupling indicated by the bond connecting different sites. In model 2*a* we consider a wire D (the "driver") carrying a current $J_D$ and investigate the possibility of current transfer to wire A through a coupling region defined by coupling matrix elements $V_{ij}$ between $N_{DA}$ pairs of neighboring atoms $i$ and $j$ on different wires. This coupling will appear in the system Hamiltonian written in the site representation. In Fig. 2a, where $N_{DA} = 3$, this coupling region includes atoms 1,2 and 3 on wire D, 5, 6 and 7 on A and the couplings $V_{1,5}, V_{2,6}, V_{3,7}$ between them. Any charge transferred from D to A will flow to the right of site 7 and to the left of site 4. We denote these currents $J_{AR}$ and $J_{AL}$, respectively. A signature of current transfer may be taken as $J_{AR} \neq J_{AL}$, or in the case of a transient (pulse) event, $\int_{-\infty}^{\infty} dt J_{AR}(t) \neq \int_{-\infty}^{\infty} dt J_{AL}(t)$; the integrals expressing the total charge transferred rightwards and leftwards through wire A. Model 2b is similar, except that the transfer $D \to A$ is mediated through a bridging wire $B$. Here $N_{DB}$, $N_{BA}$ and $N_B$ (2,2 and 1 in this example) denote respectively the number of site pairs connecting the wires D and B, the corresponding number between wires B and A and the number of B sites between these coupling regions. In this case, the signature of current transfer is similar, except that $J_{AR}$ and $J_{AL}$ now express charge transfer rates through the "acceptor" A for rightward and leftward going driving current $J_D$. Model 2c is a version of model 2b in which the driving current is explicitly seen to originate from a circular current on the donor ring, in particular its direction between sites 1 and 2 that couple to the rest of the system reflects the circular polarization of the ring current. Also in Fig. 2c we emphasize that the nature of the accepting system A is not very important in this case. The only requirement is that some signal proportional to its population is induced in the detector. The signature of current transfer is then an asymmetry in this signal under direction reversal in the driving current. It should be clear from the examples in Fig. 2 that current transfer as defined above can take place in these tight-binding models only if the driver wire $D$ is coupled to the wires A in Fig. 2a or $B$ in Figs. 2b,c by more than a single bond. Indeed, the directional information associated with the transfer is conveyed through interference between different transfer paths. This implies that thermal interactions and dephasing process may have strong effect on this process.

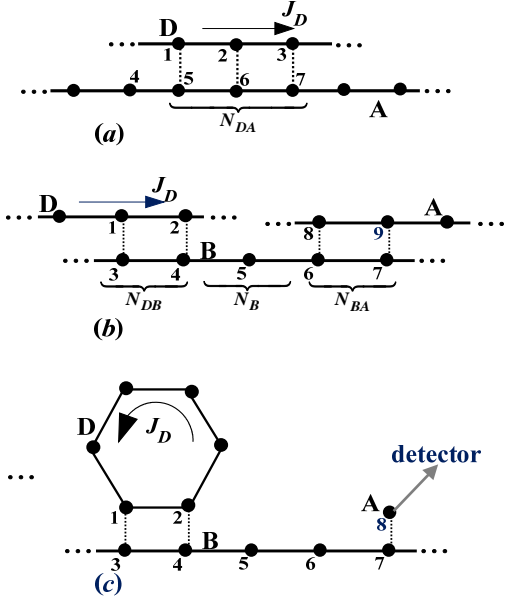

Fig. 2. Simple models of current transfer. In both models a current in wire D is transferred to wire A. In (a) the transfer is direct while in (b) it is mediated by a "bridge" wire B. Model (c) is another version of (b), where the driving current $J_D$ results from a circular current on the donor ring D.

In Ref. [1] we have described a time dependent approach to the problem of current transfer, where a transient current $J_D$ is initially generated by some pulse excitation. There are situations, e.g., those pertaining to molecular conduction phenomena, where the complementary steady state description is advantageous. The present paper presents the steady state approach to this problem. Here we obtain the steady state assumed by the system when driven by a constant current on wire D and evaluate the currents induced in other parts of the system under this driving. The current transfer property of these systems may be quantified by the assymetry factor

$$\mathcal{A} = \frac{J_{AR} - J_{AL}}{J_{AR} + J_{AL}} \qquad (1)$$

that measures the transfer of directionality information from the driving current. Our objective is to examine the dependence of this property on the molecular structure expressed by the coupling scheme, the intrachain and interchain site energies and coupling strengths and the dephasing caused by thermal interactions.

**2. Model Hamiltonian and the steady state problem**

For definiteness we focus on the model of Fig. 2b, which depicts the driving wire $D$, the bridging wire $B$ and the accepting wire A as linear tight binding chains. The corresponding Hamiltonian is $\hat{H} = \hat{H}_D + \hat{H}_A + \hat{H}_B + \hat{V}_{BA} + \hat{V}_{BD}$, where

$$H_K = \sum_{j_K \in K} E_{j_K}^{(K)} |j_K\rangle\langle j_K| + \sum_{j_K \in K} V_{j_K, j_K+1}^{(K)} |j_K\rangle\langle j_K+1| \; ; \quad K = D, A, B \tag{2}$$

and

$$V_{K,K'} = \sum_{j_K \in K, j_{K'} \in K'} V_{j_K, j_{K'}}^{(K,K')} |j_K\rangle\langle j_{K'}| \; ; \quad (K,K') = (D,B) \text{ or } (B,A) \tag{3}$$

where $\hat{H}_D$, $\hat{H}_B$, and $\hat{H}_A$ are the Hamiltonians for the D, B and A moieties, respectively and $\hat{V}^{(DB)}$, $\hat{V}^{(BA)}$ are the D-B and A-B interactions. When the driving wire is an $N_D$-member cyclic molecule as in Fig. 2c, this cyclic periodicity is reflected by the additional condition $j_D + N_D = j_D$. The magnitudes of the inter-chain coupling elements $V_{j_K, j_{K'}}^{(K,K')}$ in Eq. (3) should reflect the actual geometry of the given molecular structure, e.g. proximity between atoms belonging to different molecules. Here too we assume non-zero coupling only between sites on different wires that are nearest to each other, e.g, the site pairs (1,5), (2,6) and (3,7) in Fig. 2a, or (1,3), (2,4) in Figs 2b and 2c. In what follows we consider the particular case where each wire is a sequence of similar sites and all intersite couplings are the same for similar pairs of sites. Accordingly we denote

$$\alpha_K = E_{j_K}^{(K)}; \qquad \beta_K = V_{j_K, j_K+1}^{(K)} \; ; \quad V_{KK'} = V_{j_K, j_{K'}}^{(K,K')} \tag{4}$$

In Ref. [1] we have considered the time evolution that follows the excitation of ring current in the driving wire. If the latter is a ring of $N_D$ equivalent sites this is represented by the Bloch wavefunction

$$\Psi(t=0) = \sqrt{\frac{a}{L}} \sum_{j_D=1}^{N_D} e^{i(j_D-1)ka} |j_D\rangle. \tag{5a}$$

with

$$k = \frac{2\pi M}{aN_D}; \quad M = 0, \pm 1, ..., \pm N_D - 1; \quad L = N_D a \tag{5b}$$

where $a$ is the inter-site distance. Here the driving wire is restricted to remain in this state, and we require the steady state assumed by the rest of the system under this restriction. Obviously, the only relevant sites on the driver are those that are directly coupled to the *B* wire, e.g. sites 1, 2 in Figs 2b,c. The driving restriction in this case takes the form

$$\psi_D(t) = c_1(t)|1\rangle + c_2(t)|2\rangle = (\bar{c}_1|1\rangle + \bar{c}_2|2\rangle)e^{-(i/\hbar)Et} \quad (6a)$$

$$\bar{c}_2 = \bar{c}_1 e^{ika} \quad (6b)$$

where *E* and *k* are related by the characteristic 1–d tight binding dispersion relation

$$E = \alpha_K + 2\beta_K \cos(ka) \qquad (\text{Here } K = D) \quad (7)$$

For the system to reach steady state, the acceptor wire A has to be infinite, as in Fig. 2b. Alternatively, the population in the acceptor site A must be assumed to be depleted by coupling to some external sink as in Fig. 2c. Such depletion is not part of the Hamiltonian (2), (3) and must be added as phenomenological terms in the time dependent Schrödinger equation. In either case the ensuing currents in the acceptor are the detected outcome of the driving process.

For the model Hamiltonian of Eqs. (2), (3) the time dependent Schrödinger equation in the site representation reads

$$\hbar \frac{dc_n}{dt} = -iE_n c_n - i\sum_\alpha V_{n,\alpha} c_\alpha \quad (8)$$

where $\alpha$ goes over all sites that couple (with coupling elements $V_{n,\alpha}$) to site *n*. At steady state driven as described by Eq. (6) (or its equivalent for the system of Fig. 2a) we expect a solution of the form

$$c_n(t) = \bar{c}_n e^{-iEt/\hbar} \quad (9)$$

Using (9) in (8) leads to

$$0 = -i(E_n - E)\bar{c}_n - i\sum_\alpha V_{n,\alpha} \bar{c}_\alpha \quad (10)$$

Eqs. (10) are linear algebraic equations for the coefficients $\{\bar{c}_n, n \in B, A\}$ that define the steady state wavefunction, $\Psi(t) = \sum_{n \in B,A} c_n(t)|n\rangle$, on the B and A wires. Terms involving $\{\bar{c}_n, n \in D\}$ appear as inhomogeneous source terms in these equations.

The driving current on the D wire is a source of charge carriers in the system. For Eqs. (8) to yield the steady state form at long time, Eqs. (9),(10), it has to be supplemented by terms describing population absorption. In Ref. [1] this was achieved by assigning (real and positive) decay rates $\gamma_j$ to some sites by replacing $E_j$ by $E_j - (1/2)\gamma_j$ for these sites in Eqs. (8) and (10). The outgoing rate from the acceptor site A in Fig. 2c may be described in this way. Alternatively we may consider the infinite B and A wires of Figures 2 as exact models for effectively damping population in the relevant (observed) part of the system. In this case Eqs. (8) and (10) represent an infinite set of equations that can be made finite by the usual technique of separating the overall system into an interior "relevant" part and the remaining exterior part, and accounting for the effect of the latter on the dynamics of the former by an appropriate "self energy" term. In particular, in Fig. 2a, the effect of an exterior part defined as the infinite linear chain extending beyond the cutoff site 7 on A is manifested by modifying Eq. (10) for this site according to,

$$0 = -i\left(E_A + \Sigma_A(E) - E\right)\bar{c}_7 - i\beta_A \bar{c}_6 - iV_{AD}\bar{c}_3 \tag{11}$$

$\Sigma_A(E)$ is the self energy of a 1-dimensional tight binding wire, with real and imaginary parts $\Lambda_A(E)$ and $-i(\Gamma_A(E)/2)$ respectively,

$$\Sigma_K(E) = \frac{E - E_K - \sqrt{(E - E_K)^2 - 4\beta_K^2}}{2} \equiv \Lambda_K(E) - \frac{i}{2}\Gamma_K(E) \tag{12}$$
$$K = B, A$$

A full finite set of steady state equations for, e.g., model 2b can now be written in a straightforward manner:

$$\begin{aligned}
0 &= -i\left(E_B + \Sigma_B(E) - E\right)\bar{c}_3 - i\beta_B \bar{c}_4 - iV_{BD}\bar{c}_1 \\
0 &= -i\left(E_B - E\right)\bar{c}_4 - i\beta_B \bar{c}_3 - i\beta_B \bar{c}_5 - iV_{BD}\bar{c}_2 \\
0 &= -i\left(E_B - E\right)\bar{c}_5 - i\beta_B \bar{c}_4 - i\beta_B \bar{c}_6 \\
0 &= -i\left(E_B - E\right)\bar{c}_6 - i\beta_B \bar{c}_5 - i\beta_B \bar{c}_7 - iV_{BA}\bar{c}_8 \\
0 &= -i\left(E_B + \Sigma_B(E) - E\right)\bar{c}_7 - i\beta_B \bar{c}_6 - iV_{BA}\bar{c}_9 \\
0 &= -i\left(E_A + \Sigma_A(E) - E\right)\bar{c}_8 - i\beta_A \bar{c}_9 - iV_{AB}\bar{c}_6 \\
0 &= -i\left(E_A + \Sigma_A(E) - E\right)\bar{c}_9 - i\beta_A \bar{c}_8 - iV_{AB}\bar{c}_7
\end{aligned} \tag{13}$$

or

$$\mathbf{Mc} = \mathbf{d} \tag{14}$$

where **c** is the column vector $\text{trans}(\bar{c}_3, \bar{c}_4, \bar{c}_5, \bar{c}_6, \bar{c}_7, \bar{c}_8, \bar{c}_9)$, **M** is the matrix multiplying this vector in Eq. (13) and **d** is the driving vector $\text{trans}(iV_{BD}\bar{c}_1, iV_{BD}\bar{c}_2, 0, 0, 0, 0, 0)$ ("trans" denotes transpose). Note that in the phenomenological approach discussed above and in Ref. [1] $\Sigma_K(E); K = B, A$ are replaced by constants damping terms $-(1/2)i\gamma_j$ (for site *j*) representing coupling to some arbitrary broad-band dissipation channels. Using the self energies associated with infinite 1-dimensional chains has the advantage of providing reflection-less interfaces, making it easier to identify and quantify current transfer processes in steady-state situations.

Inverting (14) and using Eq. (6b) yields all coefficients in terms of $\bar{c}_1$. This makes it possible to evaluate all currents in the system in terms of the driving current on wire D as described in the next section.

### 3. Steady-state currents and current asymmetry factors

Using Eqs. (8)-(10) with $E_n$ replaced by $E_n + \Sigma_n(E)$ when *n* is an edge site yields the following steady state (SS) equation for the population on site *n*

$$0 = \left(\frac{d|c_n(t)|^2}{dt}\right)_{SS} = \sum_\alpha \frac{2V_{n,\alpha}}{\hbar} \text{Im}(c_\alpha c_n^*) - \frac{\Gamma_n(E)}{\hbar}|c_n|^2 \delta_{n,edge} \qquad (15)$$

where as in Eq. (8) the sum goes over all sites that couple to site *n* with coupling elements $V_{n,\alpha}$ ($V_{n,\alpha} = \beta_K$ if both *n* and *α* belong to wire K; $V_{n,\alpha} = V_{K,K'}$ if these sites are nearest neighbors belonging to different wires *K* and *K'*). The term containing $\Gamma_n(E) = -2\text{Im}\Sigma_n(E)$ contributes only if site *n* is an edge site on the bridge ($\Gamma_n = \Gamma_B$), or on the acceptor wire ($\Gamma_n = \Gamma_A$).

Eq. (15) is a continuity law that describes conservation of probability, and can be used to identify the current between any two sites as well as current going into and out of a given system. In particular, Eq. (15) implies that the current from site *n*-1 to site *n* on a given wire *K* is

$$J_{K(n-1 \to n)} = \frac{2\beta_K}{\hbar} \text{Im}(c_{n-1} c_n^*) \qquad (16)$$

and the current out of the system at the edge site *n* on wire K is

$$J_{K(n \to out)} = \frac{\Gamma_n(E)}{\hbar}|c_n|^2 \qquad (17)$$

In what follows, unless otherwise stated, we assign positive signs to currents from left to right, from D to B, from D to A and from B to A.

Before continuing with our main theme, it will prove useful to consider the application of Eqs. (16) and (17) to the free particle motion on a linear tight binding chain. Putting $c_n = c_{n-1} e^{ika}$ we find[1]

$$J_{K(n-1 \to n)} = -\frac{2\beta_K}{\hbar} |c_n|^2 \sin(ka) \qquad (18)$$

For a current in the direction $n-1 \to n \to out$, Eq. (16) implies that $\beta_K \sin(ka)$ needs to be negative. Current conservation implies that

$$\Gamma_K(E) = 2|\beta_K \sin(ka)| \qquad (19)$$

and using (cf. Eq. (7)) $\cos(ka) = (E - \alpha_K)/2\beta_K$ leads to

$$\Gamma_K(E) = 2|\beta_K| \sqrt{1 - \left(\frac{E - \alpha_K}{2\beta_K}\right)^2} \qquad (20)$$

which is consistent with Eq. (12). Furthermore Eqs. (7) and (12) imply

$$\Lambda_K(E) = \beta_K \cos(ka) = \frac{E - \alpha_K}{2} \qquad (21)$$

Eqs. (20) and (21) holds for $E$ inside the K-wire energy band.

Consider now the steady state currents induced in the system by the driving current on the D wire. In particular, for the DA model of Fig. 2a we focus on the current out of the A wire to the right and to the left. It is convenient to define both as positive quantities

$$J_A^{right} = \frac{2\beta_A}{\hbar} \text{Im}\left(c_{eright-1} c_{eright}^*\right) = \frac{\Gamma_A(E)}{\hbar} |c_{eright}|^2 \qquad (22)$$

$$J_A^{left} = \frac{2\beta_A}{\hbar} \text{Im}\left(c_{eleft+1} c_{eleft}^*\right) = \frac{\Gamma_A(E)}{\hbar} |c_{eleft}|^2 \qquad (23)$$

where *eright* and *eleft* denote respectively the right and left edge sites on the A wire. Note that by symmetry, $J[right, A, k_D] = J[left, A, -k_D]$. A non-zero current asymmetry factor

---

[1] Near the bottom of the band, $k \to 0$ (see Eq. (7)), $J_{K(n-1 \to n)} = -(2\beta_K/\hbar)|c_n|^2 ka$. The fact that for positive $k$ a positive (left to right) current is associated with negative $\beta_K$ is related to the fact that the kinetic energy operator on a grid of spacing $h$ is given by $-f''(x) \approx -h^{-2}[f(x+h) - 2f(x) + f(x-h)]$.

$$\mathcal{A}_1 = \frac{J_A^{left} - J_A^{right}}{J_A^{left} + J_A^{right}} \quad (24)$$

is a signature of current transfer, i.e. it shows that directional information is transferred together with the charge transfer process.

We can consider similar asymmetry factors for the B and A wires in the DBA model of Fig. 2b. However, a better measure in this case is the dependence of the total charge transmitted to the A wire on the direction of the driving current on the D wire, characterized by the asymmetry factor

$$\mathcal{A}_2 = \frac{J_A^{total}(-k_D) - J_A^{total}(k_D)}{J_A^{total}(-k_D) + J_A^{total}(k_D)} \quad (25)$$

$\mathcal{A}_2$ is directly related to the observations in Refs [2,3] and it quantifies the effect of the current transfer information mediated through the bridge B on the $D \to A$ charge transfer. It can be studied as function of the energy level positioning of the B wire relative to the D and A wires, and as a function of the B wire length. We will sometimes also consider the differently normalized quantities

$$\bar{\mathcal{A}}_1 = \frac{J_A^{left} - J_A^{right}}{|J_D|} \quad (26)$$

$$\bar{\mathcal{A}}_2 \equiv \frac{J_A^{total}(-k_D) - J_A^{total}(k_D)}{|J_D|} \quad (27)$$

where $J_D$ is the donor (driving) current.

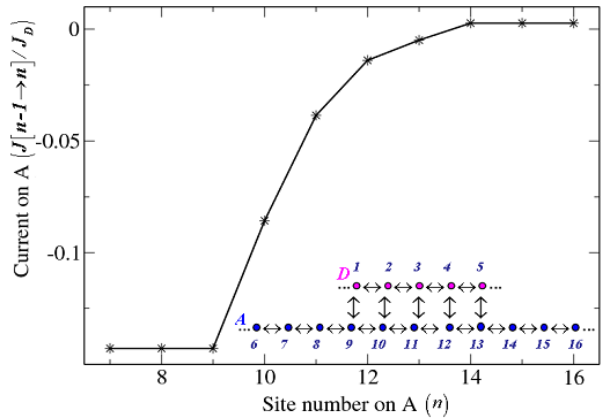

Fig. 3. Current distribution on the A wire of the DA system shown in the inset, characterized by 5-site coupling between the D and A wires. Parameters are: $|J_D|=1$, $E_D = E_A = 0$,

$\beta_D = \beta_A = 0.1$, $V_{DA} = 0.01$ and the injection energy is $E = -0.15$. The phase $k_D a = \arccos\left[(E - E_D)/2\beta_D\right]$ is taken positive, implying that the driving current goes leftward.

Fig. 3 shows the current distribution in the A wire of the DA system shown in the inset. Here and below all energy units are relative; and the reader may assign them as convenient, e.g., take all the numbers given for energies to express electron volts. The driving current induces left and right going currents at the left and right sides, respectively, of the A wire. The current is of course position independent in all parts of A that are not coupled to the driving wire, and changes in the coupling region. The larger leftward current on A reflects the directionality transfer, i.e. the current transfer character of the process. This asymmetry, and its counterpart in the DBA system are expressed in the figures below using the current asymmetry factors $\mathcal{A}_1$ and $\mathcal{A}_2$.

In what follows we show some computed results for these asymmetry factors that indicates their dependence on system parameters and structure. The later is expressed in terms of the number of links (coupled site-pairs), $N_{DA}$, connecting the D and A wires in the DA system, the corresponding numbers $N_{DB}$ and $N_{BA}$ in the DBA system and the length $N_B$ of the bridge segment separating the DB and BA coupling regions in the DBA system (see Fig. 4). In the structure displayed in Fig. 1 $N_{DA}$ and $N_{DB}$ are the number of dashed lines connecting the ring to the helical bridge on the left and right, respectively, while $N_B$ corresponds to the length of the helical bridge.

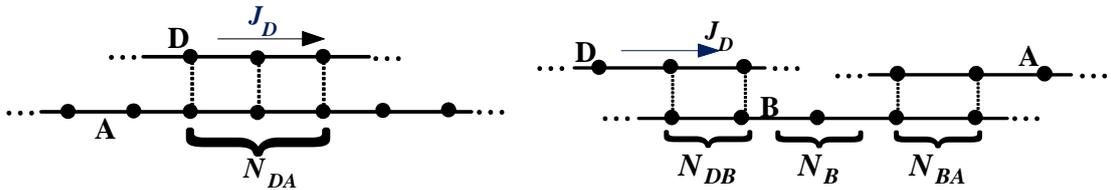

Fig. 4. Structural parameters in the DA and DBA systems. $N_{DA}$, $N_{DB}$ and $N_{BA}$ are the number of links connecting the wires D, B and A. $N_B$ corresponds to the length of the bridge segment between the D-B and the B-A coupling regions in the DBA system.

Fig. 5 shows, for the DA system, the current asymmetry factor $\mathcal{A}_1$ for the DA system of Fig. 4 as a function of the number of links (coupled site-pairs), $N_{DA}$, connecting the D and A wires. Obviously $\mathcal{A}_1 = 0$ for $N_{DA} = 1$. Fig. 6 shows the corresponding property $\mathcal{A}_2$ for the DBA system of Fig. 4. We see that in both cases asymmetry increases, then saturates near 1 (when the response current becomes nearly unidirectional), as $N_{DA}$ increases.

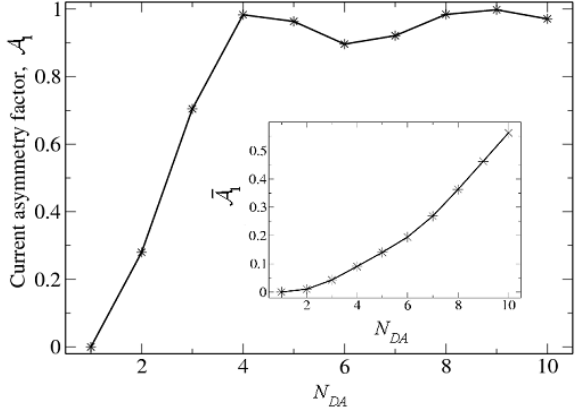

Fig. 5. The current asymmetry factor $\mathcal{A}_1$ displayed against the number of links, $N_{DA}$ connecting the D and A wires in the DA system (Fig. 4). The inset shows the same data, presented in terms of $\bar{\mathcal{A}}_1$ plotted against $N_{DA}$. Parameters are same as in Fig. 3.

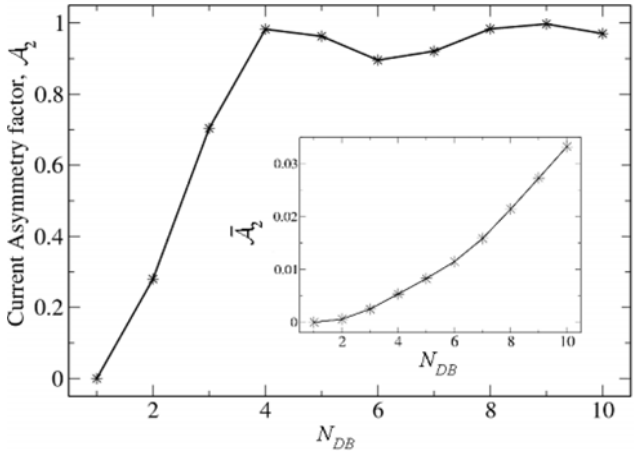

Fig. 6. The current asymmetry factor $\mathcal{A}_2$ displayed against the number of links $N_{DB}$ connecting the D and B wires in the DBA system of Fig. 4. Parameters are similar to Fig. 3: $E_D = E_B = E_A = 0$; $\beta_D = \beta_B = \beta_A = 0.1$, $V_{DB} = V_{BA} = 0.01$, $N_{BA} = 1$ and injection energy $E = -0.15$. The inset shows the same data, presented as $\bar{\mathcal{A}}_2$ plotted against $N_{DB}$.

The large asymmetry factors seen in these results indicate the current transfer character of these processes, which is seen both in the direct transfer (the DA model) and the bridge assisted (DBA model) cases. Note that the infinite extent of the bridge is important in this steady state process. For a short finite bridge the directionality may be diminished or lost by wavefunction reflection at the edge of the B wire. In this situation we indeed find much smaller asymmetry factors. It should be further noted that in this respect the steady state situation is different from the transient process considered in Ref. [1], where, for short pulses reflection does not sets in appreciably during the process lifetime.

Another important factor already discussed in Ref. [1] is the resonance or non-rsonance nature of the transfer process. The results displayed in Figs. 3, 5, 6 correspond to resonance transmission, where site energies in all wires were taken equal. The current transfer equivalent to non-resonant charge transfer, the so called superexchange mechanism, would be the DBA system where the bridge energy $E_B$ is different from the site energies $E_D = E_A$ in the "donor" and "acceptor" wires. Because of the finite bandwidths, the onset of non-resonance transfer depends on the wires band structure and on the injection energy $E$. In the calculation presented below, we use $E_D = E_A = 0$, $\beta_D = \beta_A = \beta_B = 0.1$, and an injection energy $E = -0.15$ and display the transfer behavior with respect to changing bridge energy $E_B$. Because energy bands in these tight binding wires range within $E_K \pm 2\beta_K$ (K=D, B, A), non-resonant transfer sets in as $E_B$ increases above $E_B = 0.05$ or decreases below $E_B = -0.35$. Fig. 7 shows the current transfer behavior of this system in the off-resonance regime, $E_B > 0.05$. We see a strong exponential damping of the current transfer property expressed by the asymmetry factor $\mathcal{A}_2$. It should be emphasized that, as defined, $\mathcal{A}_2$ is not sensitive to the exponential damping of the charge transfer itself, and its behavior reflects only the erasure of the current transfer property, i.e., the directionality.

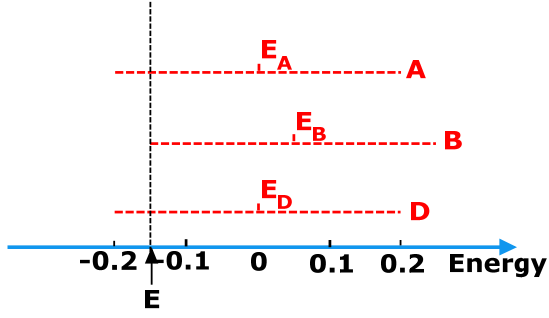

Fig. 7. A schematic illustration of the band structures in the DBA system discussed in the text. The horizontal (red) dashed lines correspond to the bands of the D, B and A wires, as indicated. E is the injection energy, which, in the situation shown, is at the lower band edge of the B wire.

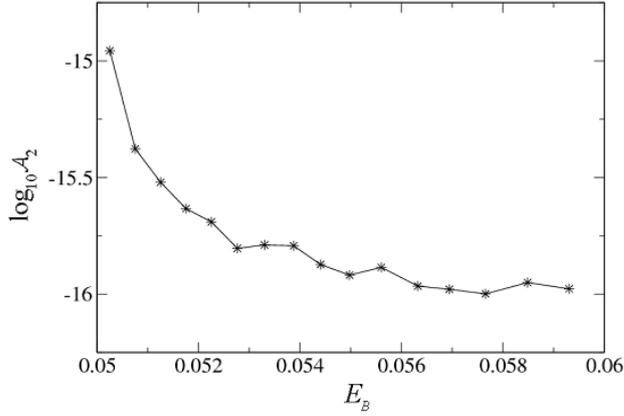

Fig. 8. The current asymmetry factor $\mathcal{A}_2$ plotted against the bridge site energy $E_B$ for the DBA system in the off-resonance regime. The parameters used in this calculation are $E_D = E_A = 0$, $\beta_D = \beta_B = \beta_A = 0.1$, $V_{DB} = V_{BA} = 0.01$, $E = -0.15$, $N_{DB} = 2$, $N_B = 1$, $N_{BA} = 1$. Shown is the current transfer behavior of the system in the non-resonance regime, $E_B > 0.05$. Coarse grained averaging was applied to reduce numerical errors that results from computing small differences between relatively large numbers.

Fig. 8 shows that the current transfer property dies out quickly as we go into the off-resonance transfer regime. One could naively expect that such a trend will be also seen as we approach the edge of the resonance transfer regime from the resonance side, $-0.35 < E_B < 0.05$ for the parameters used in Fig. 8. Fig. 9 shows that the situation is more complicated although the directionality transfer property exhibited by $\mathcal{A}_2$ indeed goes smoothly towards (essentially) zero as we approach the band edge. It is interesting to note that the charge transfer itself, expressed by the absolute current in the A wire, is singular at the band edge. This is seen in Fig. 10 that depicts the $\bar{\mathcal{A}}_2$ analog of Fig. 9. Similar behavior is obtained for the individual components, $J_A^{total}(\pm k_D)/J_D$ of $\bar{\mathcal{A}}_2$), shown in the inset. Remarkably, the individual right and left currents on the A wire can be larger than the driving current $J_D$.

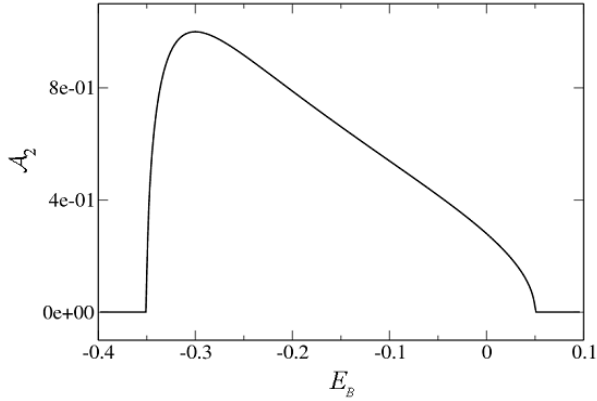

Fig. 9. The current asymmetry factor $\mathcal{A}_2$ plotted against the center band energy $E_B$ of the B wire in the DBA system. Parameters are same as in Fig. 8.

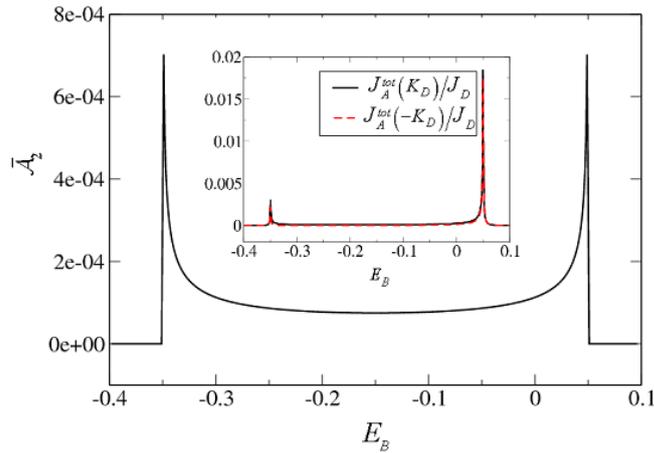

Fig. 10. Same data as in Fig. 9, now represented in terms of $\bar{\mathcal{A}}_2$, Eq. (27), plotted against $E_B$ in the resonance transmission regime. Parameters are same as in Fig. 8. The inset shows the individual contributions, $J_A^{total}(\pm k_D)/J_D$.

It is easy to understand the origin of the band-edge singularities seen in Fig. 10 by considering the simpler model of Fig. 11 (left). In this model, with a single link connecting the D and A wires, directionality information cannot be transferred. Indeed, the only property of the D wire that affects the A wire is the amplitude oscillation at site 2 determined by the injection energy $E$, $c_2(t) = \bar{c}_2 e^{-(i/\hbar)Et}$. At steady state, the amplitude $c_3(t)$ behaves similarly, $c_3(t) = \bar{c}_3 e^{-(i/\hbar)Et}$, where $\bar{c}_3$ satisfies

$$0 = -i(E_A + 2\Sigma_A(E) - E)\bar{c}_3 - iV_{AD}\bar{c}_2 \quad \text{or} \quad \bar{c}_3 = \frac{V_{AD}}{E - E_A - 2\Sigma_A(E)}\bar{c}_2 \qquad (28)$$

where $\Sigma_A(E)$ is given by Eq. (12) with $K = A$. The factor 2 multiplying $\Sigma_A(E)$ in (28) results from the fact that site 3, as part of the infinite A wire, is coupled to two identical semi-infinite parts of this wire. Using (12) in (28) leads to

$$\bar{c}_3 = \frac{V_{AD}}{i\Gamma_A(E)}\bar{c}_2 \qquad (29)$$

and from (17) it follows that the current (left or right) out of site 3 is

$$J_{(3 \to right)} = \frac{|V_{AD}|^2}{\hbar \Gamma_A(E)}|\bar{c}_2|^2 \qquad (30)$$

which diverges as $\Gamma_A(E)$ goes to zero at the band edge. It is this singular behavior that manifests itself also in the more complex situations presented above, but it is important to note that its appearance is not universal. For example, if wire A is replaced by a system of $n$ identical wires coupled to the driver site 2 via node 3 (Fig. 11, right, shows the $n = 4$ case), Eq. (28) is replaced by $\bar{c}_3 = V_{32}\bar{c}_2(E - E_A - n\Sigma_A(E))^{-1}$. The singularity at the band edge is seen to be specific to the $n = 2$ case.

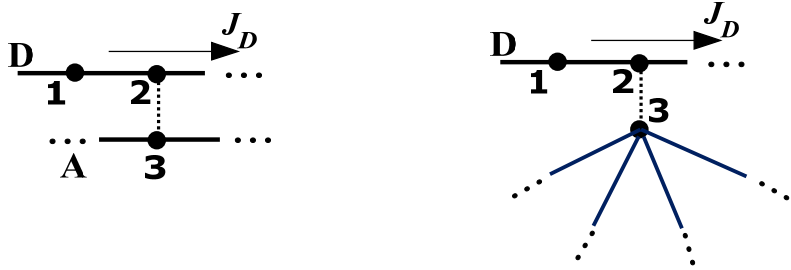

Fig. 11. Simple models used to demonstrate the origin of band-edge effects in current transfer dynamics. The model on the right is the $n=4$ analog of the ($n=2$) model on the left which in turn is a single DA link analog of the model of Fig. 2*a*.

We end this Section with some notes on this latest observation that shed some light on the current transfer formalism and its relationship to scattering theory. The results displayed in figures 3, 5-9 give information on different currents in a steady state system under the condition that a given uniform and unidirectional current flows in the driving wire. It is important to realize that these currents do not have to satisfy any continuity condition with respect to the driving current, therefore there is no contradiction in the observation that, under some conditions, a current in the A wire that is consistent with a given current in the D wire is larger than that current. In the equivalent scattering problem the system is driven by an incoming current in one of the channels (e.g. the left side of the D wire), and the transmitted currents in other channels together with the reflected current in the original channel (which by definition is missing in the present formulation) have to satisfy the usual continuity relationship that implies current conservation. In the scattering theory analog of the current transfer problem of Figs 8,9, we find[9] that when $E_B$ approaches the band edge from inside the band, the reflection coefficient becomes essentially 1, i.e. the net current on the D wire is essentially zero. The equivalent result of the current transfer calculation in which the driving current is restricted to remain constant, is divergence of the current in other channels. While this result is mathematically sound, its physical implication is that moving $E_B$ towards the singularity cannot be done while sustaining a constant current in the driving wire. We discuss this in more details in Ref. [9].

## 5. Steady state current transfer in the density matrix formalism: The effect of dephasing

In the tight binding model and in the local site representation, current transfer, the transfer of directional information in the course of a charge transfer process, arises from interference between different transfer paths. This is most directly realized by the observation that at least two intersite links between wires are needed for current transfer to be realized. It is therefore of interest to examine the effect of dephasing (decoherence) on the efficiency of these processes. We assume that motion on the driving wire D remains coherent (Bloch wave), and study the effect of dephasing on the B wire. For this purpose, we first recast the steady state approach to current transfer in the density matrix language and examine the ensuing Liouville space dynamics. Within the resulting framework we can study relaxation effects and in particular the effect of "pure" (see below) dephasing, i.e. processes that damp non-diagonal elements of the density matrix without affecting the population (diagonal elements) dynamics.

For a closed quantum mechanical system described by the Schrödinger equation $i\hbar d\Psi/dt = \hat{H}\Psi$, the transition to a Liouville space description, $i\hbar d\hat{\rho}/dt = [\hat{H}, \hat{\rho}]$ is straightforward. For our tight-binding model in the site representation explicit equations for the density matrix elements $\rho_{nm} = c_n c_m^*$ can be obtained using Eqs. (8) to get

$$d\rho_{nm}/dt = (dc_n/dt)c_m^* + c_n(dc_m^*/dt). \qquad (31)$$

Our goal is to extend these equations to steady state situations involving driving and damping as detailed above. To this end we note that from Eq. (9) it follows that at steady state $d\rho_{nm}/dt = 0$. Furthermore, population damping enters in the time evolution of diagonal density matrix elements as $d\rho_{nn}/dt = ... - \gamma_n \rho_{nn}$ and in the corresponding equations for non-diagonal elements as $d\rho_{nm}/dt = ... - (1/2)(\gamma_n + \gamma_m)\rho_{nm}$. This remains true also in steady state situations involving infinite wire systems, where apparent damping results from the imaginary part of the self energy of edge sites, as discussed in Section 3, i.e. $d\rho_{nm}/dt = 0 = ... - (1/2)(\Gamma_n(E) + \Gamma_m(E))\rho_{nm}$.

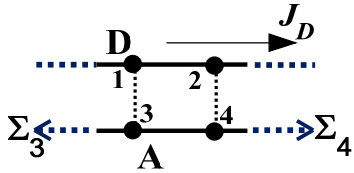

Fig. 12. A simple model used in the text to analyze current transfer in the density matrix (Liouville space) formalism.

The remaining task is then to implement the driving conditions in these steady state Liouville equations. For this purpose we follow the procedure of an early application of such ideas by Segal and Nitzan.[10, 11] In the appendix we demonstrate this procedure for the 2-link version of model 2a, which is shown in Fig. 12. It leads to Eqs. (32), (33), (34) that provide a full solution to the problem by expressing $\rho_{ij}$, $i, j \in D, A$, and the associated currents, $J[n-1 \rightarrow n] \equiv 2\left(\beta_{n,n-1}/\hbar\right) \text{Im}\,\hat{\rho}_{n-1,n}$, in the D and A wires, in terms of properties of the driving current. A generalization of the same procedure can be used in more complex situations.[12] It should be pointed out, however, that this calculation involves a certain approximation: since damping of non-diagonal matrix elements is introduced below in the site representation and not in the eigenstates basis, it does not correspond to a completely pure dephasing, i.e. it induces a small inelastic component in the outgoing electron energy, in the range $\delta E \sim \gamma$ about the injected energy $E$. Still, the electron self energy $\Sigma_K$ ($K = B, A$) is evaluated at $E$. By computing transmission using values of $\Sigma_K(E')$ with $E' = E \pm (1/2)\gamma$, we have verified that the error associated with this approximation is small for the range of dephasing rates used in the present study.

In what follows we present some examples that show the effect of dephasing, introduced as described above, on current transfer processes. In these calculations we have assigned dephasing rates $\gamma$ to the $N_{DA}$ sites on the A wire that are linked to the D wire in the DA system and to the $N_{DB} + N_B + N_{BA}$ sites on the B wire that connect between the D and A wires in the DBA system (Fig. 4). Results are shown in Figs. 13 and 14, respectively. Interestingly, while the current transfer efficiency diminishes with increasing $\gamma$, the effect persists up to relatively large values of the dephasing rate, of the order of other energetic parameters in the system. A similar observation was made in the time domain study of Ref. [1].

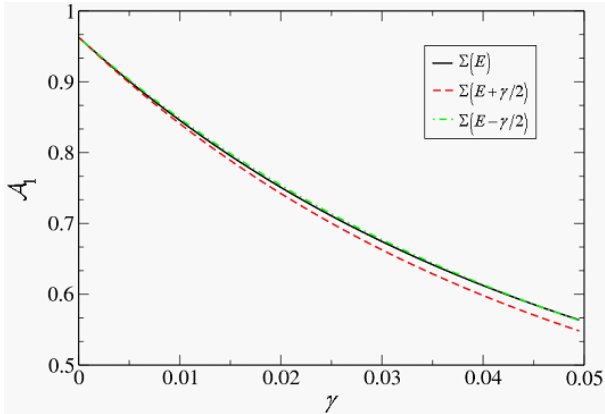

Fig. 13. The current asymmetry factor $\mathcal{A}_1$ plotted against the dephasing rate $\gamma$ for a DA system (Fig. 4) characterized by the parameters $N_{DA} = 5$, $E_D = E_A = 0$, $\beta_D = \beta_A = 0.1$, $V_{DA} = 0.01$ and $E = -0.15$. The phase $k_D a = \arccos\left(\dfrac{E - E_D}{2\beta_D}\right)$ was taken positive, implying leftward driving current. Also shown are results obtained from using $\Sigma_A(E + \gamma/2)$ (dashed line, red) and $\Sigma_A(E - \gamma/2)$ (dash-dotted line, green), instead of $\Sigma_A(E)$ in this calculation.

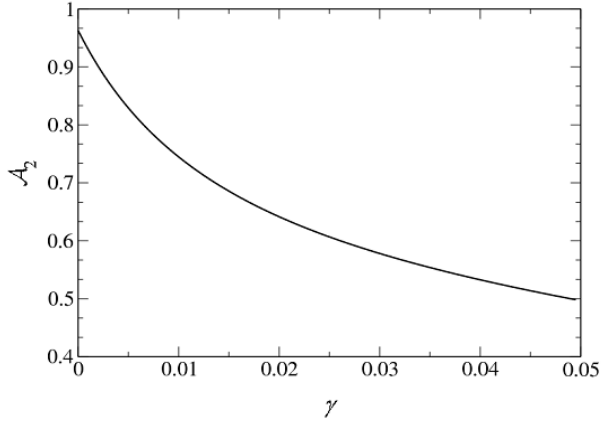

Fig. 14. The current asymmetry factor $\mathcal{A}_2$ plotted against the dephasing rate $\gamma$ for a DBA system (Fig. 4) characterized by the parameters $N_{DB} = 5$, $N_B = 3$, $N_{BA} = 1$, $E_D = E_B = E_A = 0$, $\beta_D = \beta_B = \beta_A = 0.1$, $V_{DB} = V_{BA} = 0.01$.

## 6. Conclusion

We have used the term current transfer to describe a process in which a charge carrier is transferred together with information about its motion. With the tight binding model and the local site representation used in the present work, current transfer is a coherent phenomenon resulting from interference between tunneling paths. We have investigated the dependence of this phenomenon on several key systems parameters focusing on charge transfer in two simple models, one that involves two (donor and acceptor or driving and driven) tight binding wires, and one with an additional wire that plays the role of a bridge between the two.

The present paper advances a steady state theory of current transfer, thus supplementing a previous study of this phenomenon in the time domain. On the experimental side this corresponds to the optical excitation in Refs. [2, 3] if the pulse illumination was replaced by a continuous one. Another possiblity is to attach a ring molecule to a current carrying molecular wire, mimicking a setup similar to the model in Fig. 2c but without the A site. Such a setup will work in reversal of the operation described in the discussion of Fig. 2c in Section 1, with the current in the linear wire driving a circular current in the ring, which may perhaps be detected by its magnetic field.[13]

The current transfer theory presented in this paper is based on a free electron model. Electron-electron interactions can lead to current transfer phenomena also in diffusive transport.[14] Such interactions are also believed to dominate charge fractionalization in the momentum conserving current transfer observed between parallel mesoscopic wires by Steinberg et al.[15] In our case current transfer originates from interference between different transmission or tunneling paths. Such interference phenomena in molecular wires and other nanodots have received considerable attention in recent years[16-21] We have also reformulated this theory in the density matrix representation, emphasizing the difference between the resulting steady state equations and those that could be inferred from a simplistic use of the Liouville equation. This formulation makes it possible to investigate decoherence effects on the current transfer phenomenon, demonstrating the expected reduction in current transfer efficiency in the presence of dephasing.

The current transfer models considered in this paper are formulated in terms of quantum dynamical equations with well defined "driving boundary conditions". It should be noted that such boundary conditions differ from the more familiar scattering boundary conditions. In the scattering case, the wavefunction in the D wire will consist of an incoming and a scattered components, with the latter containing reflected and transmitted parts. In contrast, in the driving problem considered here the D wire is restricted to carry a steady Bloch-wavefunction, supposedly imposed by some external driving (e.g. the cw analog of the experiments of Refs. [4], and [2,3], as discussed above), that is assumed to be insensitive to the dynamical processes in the rest of the system. We defer further discussion of the relationship between the two problems to a separate publication.

**Acknowledgements**

VbM thanks the Israel Minisry of Science for a fellowship received under the program for Progressing Women in Science. The research of AN is supported by the European Science Council (FP7 /ERC grant no. 226628), the German-Israel

Foundation, the Israel – Niedersachsen Research Fund, the US-Israel binational Science Foundation and the Israel Science Foundation. The research of S.S.S is supported by the University of Cyprus.

## Appendix

Here we implement the driving conditions described in Sections 1 and 2 in the steady state Liouville equations (31) for the model of Fig. 12, following the procedure of Segal and Nitzan.[10, 11] Using Eq. (31) together with Eqs. (8),(9) (supplemented by the self energy terms imposed on site 3 and 4 that are taken as edge sites of wire A) leads to the steady state equations

$$\begin{pmatrix} -i\Gamma_3 & 0 & -\beta_A & \beta_A \\ 0 & -i\Gamma_4 & \beta_A & -\beta_A \\ -\beta_A & \beta_A & -(i/2)(\Gamma_3+\Gamma_4)+\tilde{E}_{34} & 0 \\ \beta_A & -\beta_A & 0 & -(i/2)(\Gamma_3+\Gamma_4)+\tilde{E}_{43} \end{pmatrix} \begin{pmatrix} \rho_{33} \\ \rho_{44} \\ \rho_{34} \\ \rho_{43} \end{pmatrix} = V \begin{pmatrix} \rho_{31}-\rho_{13} \\ \rho_{42}-\rho_{24} \\ \rho_{32}-\rho_{14} \\ \rho_{23}-\rho_{41} \end{pmatrix}$$

(32a)

where $\tilde{E}_{nm} = \tilde{E}_n - \tilde{E}_m$, $\Sigma_n(E) = \Lambda_n(E) - (i/2)\Gamma_n(E)$ and $\tilde{E}_n = E_n + \Lambda_n(E)$. As before, we take $E_3 = E_4 \equiv E_A$ and $\Sigma_3(E) = \Sigma_4(E) \equiv \Sigma_A(E)$, however we keep the specific site designations here to make it easier to follow the derivation. Eq. (32a) is the standard steady state Liouville equation for the density matrix elements of the A wire and shows explicitly their dependence on density matrix elements that mix A and D sites. For the latter we employ again Eqs. (8), (9) and (31) supplemented by the self energy contributions to find

$$\begin{pmatrix} E-\tilde{E}_3-(i/2)\Gamma_3 & 0 & 0 & -\beta_A \\ 0 & E-\tilde{E}_4-(i/2)\Gamma_4 & -\beta_A & 0 \\ 0 & -\beta_A & E-\tilde{E}_3-(i/2)\Gamma_3 & 0 \\ -\beta_A & 0 & 0 & E-\tilde{E}_4-(i/2)\Gamma_4 \end{pmatrix} \begin{pmatrix} \rho_{13} \\ \rho_{24} \\ \rho_{23} \\ \rho_{14} \end{pmatrix} = V \begin{pmatrix} \rho_{11} \\ \rho_{22} \\ \rho_{21} \\ \rho_{12} \end{pmatrix}$$

(33a)

Eq. (33a) expresses the mixed DA density matrix elements, $\rho_{nm} = \rho_{mn}^*$; $n \in D, m \in A$ in terms of elements associated with the D wire only. As already noted in Ref. [10] these equations deviate from the standard Liouville equations. The reason is that in evaluating Eq. (31), the time derivatives associated with the driver coefficients are taken $\hbar dc_n/dt = -iEc_n$; $n = 1, 2$ (which expresses the driving condition) rather then derived from the system Hamiltonian. Further information about the driving enters through the explicit identification of $\rho_{nm} = c_n c_m^*$ for $n, m \in D$ which implies

$$\rho_{22} = \rho_{11}; \quad \rho_{21} = \rho_{12}^* = e^{ik_D a} \rho_{11} \tag{34}$$

The effect of pure dephasing can now be included in this dynamics using the standard phenomenological approach in which additional damping is assigned to non-diagonal elements of the density matrix: $d\rho_{ij}/dt = -\gamma_{ij}\rho_{ij}$. In the calculations reported below we assume that different local levels are affected independently by the thermal environment, whereupon $\gamma_{ij} = (1/2)(\gamma_i + \gamma_j)$, and furthermore take $\gamma_j = \gamma$ for all levels on the A and B wires (when applicable). Also, by definition, the driving dynamics is assumed unaffected by the thermal environment. This implies that zero dephasing should be assigned to levels in the D wire, i.e. $\gamma_j = 0$ for $j \in D$. Eqs. (32a) and (33a) then become

$$\begin{pmatrix} -i\Gamma_3 & 0 & -\beta_A & \beta_A \\ 0 & -i\Gamma_4 & \beta_A & -\beta_A \\ -\beta_A & \beta_A & -(i/2)(\Gamma_3+\Gamma_4+2\gamma)+\tilde{E}_{34} & 0 \\ \beta_A & -\beta_A & 0 & -(i/2)(\Gamma_3+\Gamma_4+2\gamma)+\tilde{E}_{43} \end{pmatrix} \begin{pmatrix} \rho_{33} \\ \rho_{44} \\ \rho_{34} \\ \rho_{43} \end{pmatrix} = V \begin{pmatrix} \rho_{31}-\rho_{13} \\ \rho_{42}-\rho_{24} \\ \rho_{32}-\rho_{14} \\ \rho_{23}-\rho_{41} \end{pmatrix} \tag{32b}$$

$$\begin{pmatrix} E-\tilde{E}_3-\frac{i}{2}(\Gamma_3+\gamma) & 0 & 0 & -\beta_A \\ 0 & E-\tilde{E}_4-\frac{i}{2}(\Gamma_4+\gamma) & -\beta_A & 0 \\ 0 & -\beta_A & E-\tilde{E}_3-\frac{i}{2}(\Gamma_3+\gamma) & 0 \\ -\beta_A & 0 & 0 & E-\tilde{E}_4-\frac{i}{2}(\Gamma_4+\gamma) \end{pmatrix} \begin{pmatrix} \rho_{13} \\ \rho_{24} \\ \rho_{23} \\ \rho_{14} \end{pmatrix} = V \begin{pmatrix} \rho_{11} \\ \rho_{22} \\ \rho_{21} \\ \rho_{12} \end{pmatrix}$$

$$\tag{33b}$$

Finally, it is interesting to note that the solutions to Eq. (33a), e.g.

$$\rho_{13} = \frac{\rho_{11} + \beta_A \rho_{12}}{X_3 + \beta_A^2/X_4}; \quad \rho_{23} = \frac{\rho_{21} + \beta_A \rho_{22}}{X_3 + \beta_A^2/X_4} \tag{35}$$

where $X_n = E - \tilde{E}_n - (i/2)\Gamma_n;\ n \in A$, satisfy

$$\frac{\rho_{13}}{\rho_{23}} = \frac{\rho_{11} + \beta_A \rho_{12}}{\rho_{21} + \beta_A \rho_{22}} = e^{-ik_D a} \tag{36}$$

This clearly remains true also for Eq. (33b), where $\Gamma_n$ is replaced by $\Gamma_n + \gamma$. In systems with more links between the D and A wires we find similarly

$$\frac{\rho_{j'n}}{\rho_{jn}} = e^{-ik_D(j-j')a} \; ; \quad n \in A; \quad j, j' \in D \tag{37}$$

i.e., these ratios behave as if $\rho_{jn} = c_j c_n^*$ also in the general case involving damping.

These relationships can be used, in more complex model, to reduce the number of equations that need to be solved, i.e. the size of matrices to be inverted.